\documentclass[12pt]{article}
\usepackage{enumerate}
\usepackage{amsfonts}
\usepackage{amssymb}
\usepackage{onlyamsmath}
\usepackage[OT4]{fontenc}
\usepackage[cp1250]{inputenc}
\title{$N = \frac{1}{2}$ Global SUSY: $R$-Matrix Approach} 
\author{ C. Gonera\footnote{e-mail: cgonera@uni.lodz.pl}  \quad M. Wodzisławski
\\ \\
Department of Theoretical Physics and Computer Science,\\
University of Łódź,\\
Pomorska 149/153, 90-236 Łódź, Poland }
\date{}

\begin{document}
\maketitle
\begin{abstract}
R-matrix method is used to construct supersymmetric extensions of  $\theta$ - Euclidean group preserving $N = \frac{1}{2}$ supersymmetry and its three- parameter generalization. These quantum symmetry supergroups can be considered as global counterparts of appropriately twisted Euclidean superalgebras. The corresponding generalized global symmetry transformations act on deformed superspaces as the usual ones do on undeformed spaces. However, they depend on non(anti)commuting parameters satisfying (anti)commutation relations defined by relevant R matrix.      
    
\end{abstract}
\section{Introduction}

Semi-classical arguments of  relativistic quantum theory referring to the occurrence of annihilation/creation processes or gravitational collapse \cite{a1}, \cite{a2}, \cite{a3} (if energy density in tiny volume is large enough) on one side and the results of string theory  in a non-trivial graviphoton  background \cite{a4} - \cite{a7} on the other, strongly suggest that a sort of non-(anti)commutative deformations (quantizations) of (super)spaces is a generic feature of quantum theories consistent with relativity \cite{a8}, \cite{a9}, \cite{a10}. However, these deformations break, at least partially, the original (super)symmetry \cite{a11} -  \cite{a18}. So, if (super)symmetry transformations are required to preserve (anti)commutation relations defining (super)space deformations they have to be reformulated somehow \cite{a19} - \cite{b21}.

A nice modification of infinitesimal (super)symmetry transformations (inspired by quantum group theory \cite{a22} - \cite{a28} ) has been proposed in the framework of so called twisted Euclidean/Poincare (super)algebras \cite{a29} - \cite{a41}. In this approach only the (super)coalgebraic sector of original (super)symmmetry algebra is modified while the (super)algebraic one is left unchanged. Thanks to that the representation content of twisted (super)symmetry is the same as that of standard one.
 
A global counterparts of twisted algebras have been considered in \cite{a42} -  \cite{a44.1}, \cite{a32}. In particular, in Ref. \cite{a44.1}, letting the original global symmetry transformation parameters to be non(anti)commutative ones and making some additional assumptions, both simple and natural from "physical" point of view, allowed us to construct generalized global transformations which preserve wide class of (anti)commutation relations defining various Euclidean superspace quantizations. Requiring these relations to be consistent with Jacobi identities leads, in particular, to global transformations which preserve so called $N = \frac{1}{2}$ supersymmetry \cite{a45} as well as its three-parameter extension \cite{a36}, \cite{a40}. The coalgebraic sector of these transformations appears to be compatible with the algebraic one. Therefore, they generate quantum symmetry supergroup \cite{a46} - \cite{a53} - a supersymmetric extension of $\vartheta$ - Euclidean group (i.e. Euclidean counterpart of $\vartheta$ -Poincare group \cite{a42}, \cite{a43}, \cite{a44}). This quantum supergroup can be considered as global counterpart of appropriately twisted Euclidean superalgebra which has been shown to preserve $ N = \frac{1}{2}$ supersymmetry \cite{a36}, \cite{a38}, \cite{a39}, \cite{a32}.

In the present paper (which can be viewed as the companion one to the Ref. \cite{a44.1}) it is explained how these  supersymmetric extensions of $\vartheta$ - Euclidean group can be constructed within the so called 
R - matrix formalism or Faddeev - Reshetikhin - Takhtajan (FRT) method \cite{a28}. In this approach the key role is played by R/B matrix which controls the  
(anti)commutation rules both for deformed superspace coordinates as well as transformation parameters of the corresponding quantum symmetry supergroup \cite{a49}. Unfortunately, in general, there is no universal algorithm allowing to construct such R/B matrix and what is more, it is even not known if the universal R matrix exists for every consistent (super)space deformation described by certain non(anti)commutative Hopf (super)algebra. Nevertheless, the case of $ N = \frac{1}{2}$ supersymmetry and its extensions belong to the class of superspace deformations described by twisted superalgebras. Then, there exists the general way of constructing  R matrix out of the appropriate twist operator \cite{a22}. Actually, as it is shown below, the R matrices describing $ N = \frac{1}{2}$ supersymmetry and its extension can also be constructed directly, starting basically with (anti)commutation relations defining superspace quantization (without referring to the twisting procedure).    
Indeed, using a vector realization $z^{A}$ of the superspace and corresponding supermatrix representation $T^{A}_{B}$ of Euclidean supergroup the relevant R matrix can be constructed by taking into account that:\\
a)(anti)commutation relations defining superspace deformation do not determine R matrix uniquely,\\
b) this gauge-like freedom can be fixed (due to a specific structure of some $T^{A}_{B}$ supermatrix elements ) if one requires supersymmetric version of RTT equation to provide consistent (anti)commutation relations for all $T^{A}_{B}$ elements.

The paper is organized as follows. Section 2 introduces supermatrix representation of $ N = 1 $ Euclidean supergroup as well as some notation and conventions, concerning supersymmetry ( we mostly follow \cite{a54}, \cite{a55}, \cite{a56}, \cite{a63}). In section 3 the main points of supersymmetric version of FRT method are reminded and the way it can be used to find R matrices controlling
algebraic sector of global symmetries of quantized  Euclidean superspaces is explained. This is illustrated by explicit construction of algebraic sector of the global transformations preserving $ N = \frac{1}{2}$ supersymmetry and its extensions. Finally, the coalgebraic structure (compatible with the algebraic one) is introduced. That completes the construction of supersymmetric extensions of $\theta $- Euclidean group. Section 4 discusses 
the global symmetries considered in section 3 ( as well as in Ref.\cite{a44.1} ) from the point of view of star products \cite{a57} - 
\cite{a62} and twisted superalgebras. In particular, it is argued that R matrices found in section 3 provide supermatrix realization of universal R - matrix constructed out of the twist operator defining relevant twisted symmetry superalgebra. In appendix some basic facts concerning supermatrices \cite{a63} can be found.\\
In order to make the paper more readable we will generally omit the prefix "anti" and use the words "commutators", "commutation rules", etc. irrespectively of the parity of variables under consideration. However, we keep the standard notation for the commutators ($[. , .]$) and the anticommutators ($\{. , .\}$).

\section{Supermatrix representation of $ N = 1$ Euclidean supergroup }

In order to apply FRT method it is convenient to introduce supervector realization of N=1 Euclidean superspace and supermatrix representation of $N=1$ Euclidean supergroup acting there. Four bosonic coordinates $x^{m},  m= 0,1,2,3$ 
and four grassmannnian ones $\eta^{\alpha}$,  $\bar{\eta}^{\dot{\alpha}}$, $\alpha =1,2$, $\dot{\alpha} = \dot{1}, \dot{2}$ form a supervector $z^{A}$

\begin{align}\label{42}
(x^{m}, \eta^{\alpha}, \bar{\eta}^{\dot{\alpha}})\longmapsto (z^{A})=\left(\begin{array}{ccc} x^{m}\\1\\ \bar{\eta}^{\dot{\alpha}}\\ \eta^{\alpha}\end{array}\right)
\end{align}

The elements $g(\xi, \bar{\xi}, a, \omega)$ of $N = 1$ Euclidean supergroup are parameterized by grassmnnian coordinates 
$\xi^{\alpha},  \bar{\xi}^{\dot{\alpha}}$, $\alpha = 1, 2,$  $\dot{\alpha}\,=\,\dot{1},\dot{2}$ (corresponding to supertranslations generators 
$Q_{\alpha}$, $\bar{Q}_{\dot{\alpha}}$  ) 
and bosonic coordinates $a^{m},\ \omega^{mn}=-\omega^{nm}$, $ m, n = 0, 1, 2, 3$ 
(corresponding to the ordinary translations $P_{m}$ and 4D-rotations $M_{mn}$, respectively). They can be represented by a supermatrix ( see Appendix ) $(T^{A}_{\  \ B})\equiv (T^{A}_{\  \ B}(\xi, \bar{\xi}, a, \omega))$

\begin{align}\label{43}
(T^{A}_{\   B})= \begin{bmatrix}(e ^{-\omega})^{m}_{\;\;n}& &a^{m} & & -i\xi^{\alpha}(\sigma^{m})_{\alpha \dot{\alpha}}
(B^{\dagger}(\omega))^{\dot{\alpha}}_{\    \dot{\beta}} & & -i\bar{\xi}^{\dot{\alpha}}(\sigma^{m})_{\alpha \dot{\alpha}}
(A^{T}(\omega))^{\alpha}_{\    \beta} \cr  & &\cr 0 &  & 1 &   &  0 &   &0&  \cr    & &\cr 0 &  &\bar{\xi}^{\dot{\rho}}& 
 & (B^{\dagger}(\omega))^{\dot{\rho}}_{\  \dot{\beta}} &  & 0 \cr \cr 0 & &\xi^ {\rho} & & 0  & 
& (A^{T}(\omega))^{\rho}_{\   \beta}     \cr  \end{bmatrix}
\end{align}\\

In the above equation  we used the standard notation for Pauli matrices: $\sigma_{m} =(iI,\sigma_{i})$ and $\bar{\sigma}_{m} =(iI, -\sigma_{i})$.

The generators of $D^{(\frac{1}{2}, 0)}$  and 
 $D^{(0,\frac{1}{2})} $ representations of $SU(2) \times SU(2)$  group (universal covering of 4D-rotation group ) read

\begin{align}\label{43a}
\sigma_{mn} = \frac{i}{4}(\sigma_{m}\bar{\sigma}_{n}-\sigma_{n}\bar{\sigma}_{m})\nonumber \\
\bar{\sigma}_{mn} = \frac{i}{4}(\bar{\sigma}_{m}\sigma_{n}-\bar{\sigma}_{n}\sigma_{m})\nonumber\\ 
\end{align}

which allows us to write the matrices  A and B, entering eq.(\ref{43}) in the form 

\begin{align}\label{3a}
&A(\omega) =e^{-\frac{1}{2}\omega^{mn}\sigma_{mn}}
&B^{\dagger}(\omega) =e^{\frac{1}{2}\omega^{mn}\bar{\sigma}_{mn}}
\end{align} 
Note that$A(\omega)$ and $B^{\dagger}(\omega)$ satisfy $A(\omega)\sigma^{m}B^{\dagger}(\omega) =(e^{-\omega})^{m}_{\   n}\sigma^{n} $ .

The $T^{A}_{\   B}$ matrices define the representation of $N=1$ Euclidean supergroup, that is

\begin{align}\label{44}
T^{A}_{\  \ C}(\xi, \bar{\xi}, a, \omega)T^{C}_{\  \ B}(\eta, \bar{\eta}, b, \omega')=T^{A}_{\  \ B}(\Lambda , \bar{\Lambda}, c, \omega'').
\end{align}
In the eq.(\ref{44}) the parameters $\Lambda,  \bar{\Lambda},  c,  \omega''$ are given by the composition rule in N=1 Euclidean supergroup 
( see \cite{a44.1} for example) :

\begin{align}\label{3}
g(\xi, \bar{\xi}, a, \omega)g(\eta, \bar{\eta},b,\omega')=& g(\Lambda, \bar{\Lambda}, c, \omega'') \\
\Lambda^{\alpha} =& \xi^{\alpha} + (A^{T}(\omega))^{\alpha}_{\   \beta}\eta^{\beta}\nonumber \\  
\bar{\Lambda}^{\dot{\alpha}} =&  \bar{\xi}^{\dot{\alpha}} + (B^{\dagger}(\omega))^{\dot{\alpha}}_{\  \dot{\beta}}\bar{\eta}^{\dot{\beta}}\nonumber \\
c^{m} =&  e(^{-\omega})^{m}_{\  n}b^{n} + a^{m} -i\xi^{\alpha}(\sigma^{m})_{\alpha \dot{\alpha}}
(B^{\dagger}(\omega))^{\dot{\alpha}}_{\   \dot{\beta}}\bar{\eta}^{\dot{\beta}}\nonumber \\
& -i\bar{\xi}^{\dot{\alpha}}(\bar{\sigma}^{m})_{\dot{\alpha} \alpha}(A^{T}(\omega))^{\alpha}_{\   \beta}\eta^{\beta}\nonumber\\
e^{\frac{i}{2}\omega_{ab}M^{ab}}e^{\frac{i}{2}\omega'_{ab}M^{ab}} =&  e^{\frac{i}{2}\omega''_{ab}M^{ab}}\nonumber 
\end{align}

Let $\left|T^{A}_{\  \ B}\right|$ be the parity of matrix element $T^{A}_{\  \ B}$. Defining the parities $\left|A\right|$ of rows and columns in such a way that $\left|A\right| = 0 $ if $ A = m, 4$ and $\left|A\right| = 1 $ if $ A = \alpha, \dot{\alpha}$ we find

\begin{align}\label{45}
\left|T^{A}_{\  \ B}\right| = \left|A\right|+\left|B\right|, 
\end{align}
i.e. $(T^{A}_{\  \ B})$ is an even supermatrix ( see Appendix)

Supermatrices representing supertranslation generators $Q_{\alpha}$, $\bar{Q}_{\dot{\alpha}}$ 
and usual translation generators $P_{m}$ read:  \\

\begin{align}\label{47}
&[Q_{\alpha}]^{A}_{\, \ B} = -(\sigma^{m})_{\alpha \dot{\alpha}}\delta^{A}_{m}\delta^{\dot{\alpha}}_{B} + i\delta^{A}_{\alpha}\delta^{4}_{B}\\
&[\bar{Q}_{\dot{\alpha}}]^{A}_{\, \ B} = 
(\sigma^{m})_{\alpha \dot{\alpha}}\delta^{A}_{m}\delta^{\alpha}_{B} - i\delta^{A}_{\dot{\alpha}}\delta^{4}_{B},\nonumber\\
&[P_{m}]^{A}_{\, \ B} = i\delta^{A}_{m}\delta^{4}_{B}.\nonumber
\end{align}
$[P_{m}]^{A}_{\, \ B}$ are even supermatrices,\\
\begin{align}
&\left|[P_{\mu}]^{A}_{\, \ B}\right| = \left|A\right| + \left|B\right|,
\end{align}
whereas  $ [Q_{\alpha}]^{A}_{\, \ B}$ and $[\bar{Q}_{\dot{\alpha}}]^{A}_{\, \ B}$ are odd ones,\\
\begin{align}
\left|[Q_{\alpha}]^{A}_{\, \ B}\right|=\left|[\bar{Q}_{\dot{\alpha}}]^{A}_{\, \ B}\right|=\left|A\right| + \left|B\right|+1.
\end{align}
Direct computation shows that the above generators verify relevant commutation relations :
\begin{align}\label{48}
(\left\{Q_{\alpha},\bar{Q} _{\dot{\beta}}\right\})^{A}_{\, \ B}&\equiv(Q_{\alpha})^{A}_{\, \ C}(\bar{Q} _{\dot{\beta}})^{C}_{\, \ B} + (\bar{Q} _{\dot{\beta}})^{A}_{\, \ C}(Q_{\alpha})^{C}_{\, \ B}\\& = 2(\sigma^{m})_{\alpha \dot{\beta}}(P_{m})^{A}_{\   \ B}\nonumber\\
(\left\{Q_{\alpha},Q _{\beta}\right\})^{A}_{\, \ B}&=0\nonumber\\
(\left\{\bar{Q}_{\dot{\alpha}},\bar{Q} _{\dot{\beta}}\right\})^{A}_{\, \ B}&=0\nonumber \\
([P_{m}, Q_{\alpha}])^{A}_{\, \ B}&=0\nonumber \\
([P_{m}, \bar{Q}_{\dot{\alpha}}])^{A}_{\, \ B}&=0.\nonumber
\end{align}

In addition one has:

\begin{align}\label{49}
(Q_{\alpha}Q_{\beta})^{A}_{\, \ B}=0\\
(\bar{Q}_{\dot{\alpha}}\bar{Q}_{\dot{\beta}})^{A}_{\, \ B}=0\nonumber \\
(P_{m}Q_{\beta})^{A}_{\, \ B}=0\nonumber \\
(P_{m}\bar{Q}_{\dot{\beta}})^{A}_{\, \ B}=0\nonumber \\
(P_{m}P_{n})^{A}_{\, \ B}=0.\nonumber
\end{align}

\section{R-matrix construction of global $ N = \frac{1}{2}$ supersymmetries } 

Replacing the commutative coordinates $z^{A}$ 
by quantities $\hat{z}^{A}$ (satisfying commutation relations defining deformed superspaces) one obtains vector representations of deformed Euclidean superspaces. In FRT approach the relations satisfied by $\hat{z}^{A}$ as well as symmetries of these relations
(basically described by non-commutative counterparts of supermatrix  $T^{A}_{\   B}$ elements) are determined by $\mathcal{B}^{AB}_{\  \ MN}$ 
matrix. It is defined in such a way that the following conditions are met \cite{a28},\cite{a48},\cite{a49}: \\  

 \begin{enumerate}[1.]
\item Equations
\begin{align}\label{50}
\hat{z}^{A}\hat{z}^{B} = \mathcal{B}^{AB}_{\  \ MN}\hat{z}^{M}\hat{z}^{N}
\end{align}\\
 reproduce the original commutation relations for  $\hat {x}^{m},\hat { \eta}^{\alpha},\hat { \bar{\eta}}^{\dot{\alpha}}$ .\\
\item Supersymmetric version of FRT equations:\\ \\ 
\begin{align}\label{51}
(-1)^{|E|(|N| +|F|)}\mathcal{B}^{AB}_{\   MN} T^{M}_{\   E}T^{N}_{\    F}=(-1)^{|M|(|B|+|N|)}T^{A}_{\   M}T^{B}_{\   N}\mathcal{B}^{MN}_{\   EF}
\end{align}\\
is consistent and gives commutation relations for $\hat{\xi}, \hat{\bar{\xi}}, \hat{a}, \hat{\omega}$ - the noncommmutative
 counterparts of Euclidean supertransformation parameters. These relations define the algebraic sector of global symmetries of deformed superspaces.
It is worth noting that the equations (\ref{51}) provide sufficient conditions for the covariance of eqs. (\ref{50}) under the generalized Euclidean 
transformations \\
\begin{align}\label{52}
\hat{z}^{A} \mapsto \hat{z}'^{A} =T^{A}_{\   B}\hat{z}^{B},
\end{align}\\
\item $\mathcal{B}$ matrix verifies Yang-Baxter equation\\ \\
\begin{align}\label{53}
(I \otimes \mathcal{B})(\mathcal{B} \otimes I)(I \otimes \mathcal{B})=(\mathcal{B} \otimes I)(I \otimes \mathcal{B})(\mathcal{B} \otimes I),
\end{align}\\
which guarantees associativity of composition rule.\\
\end{enumerate}

The above description of FRT method is somehow informal. However, it can be made mathematically sound (in the spirit of original paper \cite{a28} ) if needed.\\
Let us remark that eqs. (\ref{50}), (\ref{51}), (\ref{53}) can be equivalently written in terms of the  $\mathcal{R}$ matrix defined by: \\
\begin{align}\label{53a}
\mathcal{B}^{AB}_{\  MN} =(\mathcal{R}\sigma)^{AB}_{\  MN} = (-1)^{\left|M\right|\left|N\right|}\mathcal{R}^{AB}_{\  NM},
\end{align}

\begin{align*}
\sigma^{AB}_{\  MN} = (-1)^{\left|A\right|\left|B\right|}\delta^{B}_{M}\delta^{A}_{N}.
\end{align*}
Then the Yang-Baxter eq. (\ref{53}) reads:
\begin{align}\label{53b}
\mathcal{R}_{12}\mathcal{R}_{13}\mathcal{R}_{23} = \mathcal{R}_{23}\mathcal{R}_{13}\mathcal{R}_{12},
\end{align}
with standard notation\\
\begin{align}
&\mathcal{R}_{12}=\mathcal{R} \otimes I\\
&\mathcal{R}_{23}=I \otimes \mathcal{R}\nonumber \\
&\mathcal{R}_{13}=\Sigma_{\alpha}\mathcal{R}_{\alpha} \otimes I \otimes  \mathcal{R}_{\alpha}.\nonumber
\end{align}

It should be stressed that:
\begin{enumerate}[a)]
\item The matrices satisfying eq.(\ref{50}) do not have to fulfil automatically eqs.(\ref{51}) and (\ref{53}) . 
In particular, a matrix $\tilde{\mathcal{B}}$ verifying eq.(\ref{50}) can lead via eq.(\ref{51}) to non-consistent commutation relations for group parameters. Indeed, it appears that different results can be obtained when commutation relations including "composed"  
$T^{A}_{\   B}$ elements i.e. elements: 

\begin{align}\label{53q}
T^{m}_{ \  \dot{\gamma}} = -iT^{\alpha}_{\  \ 4}(\sigma^{m})_{\alpha \dot{\beta}}T^{\dot{\beta}}_{\   \dot{\gamma}}\\
T^{m}_{ \  \gamma} = -iT^{\dot{\alpha}}_{\  \ 4}(\bar{\sigma}^{m})_{\dot{\alpha} \beta}T^{\beta}_{\  \gamma},\nonumber
\end{align}

are computed in two ways :\\

\begin{enumerate}[1)]
\item directly using eq. (\ref{51}) or\\
\item referring to commutation relations ( determined by eq. (\ref{51})) for "simple" elements   
 $T^{\alpha}_{\  \ 4}$, $T^{\dot{\beta}}_{\   \dot{\gamma}}$, $T^{\dot{\alpha}}_{\  \ 4}$, $T^{\beta}_{\  \gamma}$ entering the expressions for composed ones.
\end{enumerate}

\item Equation (\ref{50}) defines $\mathcal{B}$ matrix up to a matrix $b^{AB}_{\  \ MN}$ which satisfies the following condition:\\

\begin{align}\label{55}
b^{AB}_{\  \ MN}\hat{z}^{M}\hat{z}^{N} = 0.
\end{align}

So, if $\tilde{\mathcal{B}}$ matrix verifies eq.(\ref{50}) then the matrix:

\begin{align}\label{54}
\mathcal{B}^{AB}_{\  \ MN} = \tilde{\mathcal{B}}^{AB}_{\  \ MN} + b^{AB}_{\  \ MN}
\end{align}

also does it.

\end{enumerate}

It appears that, in some cases (including those discussed in literature) this " gauge-like " symmetry of
  $\tilde{\mathcal{B}}^{AB}_{\  \ MN}$ matrix and the structure of $T^{A}_{\   \ B}$ matrix elements
(see eq.(\ref{53q})) can be used to construct new $\mathcal{B}$ matrices verifying all three conditions (\ref{50}), (\ref{51}), (\ref{53}).
In particular, this can be done for  $N = \frac{1}{2}$ Euclidean supersymmetry \cite{a45} and its three-parameter extensions \cite{a36} as it is
demonstrated in some detail below.\\
 
We start with $N = \frac{1}{2}$ Euclidean superspace (studied for instance in \cite{a32}, \cite{a45}, \cite{a36}, \cite{a37}, \cite{a38}, \cite{a39}, \cite{a40} ) defined by the following non-vanishing commutators: 

\begin{align}\label{56}
\{\hat{\eta}^{\alpha},\hat{\eta}^{\beta}\}&=C^{\alpha \beta} \\
[\hat{x}^{m}, \hat{\eta}^{\beta}]&= iC^{\alpha \beta}(\sigma^{m})_{\alpha \dot{\alpha}}\hat{\bar{\eta}}^{\dot{\alpha}}\nonumber \\
[\hat{x}^{m}, \hat{x}^{n}]&= \frac{1}{2}C^{\alpha \beta}(\sigma^{m})_{\alpha \dot{\alpha}}(\sigma^{n})_{\beta \dot{\beta}}[\hat{\bar{\eta}}^{\dot{\alpha}},\hat{\bar{\eta}}^{\dot{\beta}}]=C^{\alpha \beta}(\sigma^{m})_{\alpha \dot{\alpha}}(\sigma^{n})_{\beta \dot{\beta}}\hat{\bar{\eta}}^{\dot{\alpha}}\hat{\bar{\eta}}^{\dot{\beta}}\nonumber
\end{align}

It is a particular case of general Euclidean superspace deformation considered in Ref. \cite{a44.1} and corresponding to the following choice
of structure constants introduced there:
 
\begin{align}\label{55a}
C^{\alpha \beta} \neq 0&  &D^{\dot{\alpha} \dot{\beta}} =0& &E^{\alpha \dot{\beta}}=0\\
\Pi^{m \alpha}=0& &\Delta^{m \dot{\alpha}} =0& &\Theta^{mn} =0\nonumber
\end{align}

It is not difficult to find $\tilde{\mathcal{B}}^{AB}_{\ MN}$ matrix which reproduces ( via eq. (\ref{50}) ) the above
commutation relations (\ref{56}). Indeed, taking for example $\tilde{\mathcal{B}}^{AB}_{\ MN}$ matrix in the form :\\

\begin{align}\label{57}
\tilde{\mathcal{B}}^{AB}_{\  \ MN} =&(-1)^{\left|A\right|\left|B\right|}\delta^{A}_{N}\delta^{B}_{M} 
+ C^{\alpha \beta}\delta^{A}_{\alpha}\delta^{B}_{\beta}\delta^{4}_{M}\delta^{4}_{N}\\
& + C^{\alpha \beta}(\sigma^{m})_{\alpha \dot{\alpha}}(\sigma^{n})_{\beta \dot{\beta}}
\delta^{A}_{m}\delta^{B}_{n} \delta^{\dot{\alpha}}_{M}\delta^{\dot{\beta}}_{N} \nonumber \\
&+ \frac{i}{2}C^{\alpha \beta}(\sigma^{m})_{\alpha \dot{\alpha}}(\delta^{A}_{m}\delta^{B}_{\beta}\delta^{\dot{\alpha}}_{M}\delta^{4}_{N}-\delta^{A}_{\beta}\delta^{B}_{m}\delta^{4}_{M}\delta^{\dot{\alpha}}_{N})\nonumber \\&+ \frac{i}{2}C^{\alpha \beta}(\sigma^{m})_{\alpha \dot{\alpha}}(\delta^{A}_{m}\delta^{B}_{\beta}\delta^{4}_{M}\delta^{\dot{\alpha}}_{N}-\delta^{A}_{\beta}\delta^{B}_{m}\delta^{\dot{\alpha}}_{M}\delta^{4}_{N})\nonumber
\end{align}

leads to the following commutation relations for $\hat{z}^{A}$ variables:\\

\begin{align}\label{58}
\hat{z}^{A}\hat{z}^{B}-(-1)^{|A||B|}\hat{z}^{B}\hat{z}^{A}=
&C^{\alpha \beta}(\sigma^{m})_{\alpha \dot{\alpha}}(\sigma^{n})_{\beta \dot{\beta}}\delta^{A}_{m}\delta^{B}_{n}
\delta^{\dot{\alpha}}_{M}\delta^{\dot{\beta}}_{N}\hat{z}^{M}\hat{z}^{N}\\ 
&+iC^{\alpha \beta}(\sigma^{m})_{\alpha \dot{\alpha}}
(\delta^{A}_{m}\delta^{B}_{\beta}-\delta^{A}_{\beta}\delta^{B}_{m})\delta^{\dot{\alpha}}_{M}\hat{z}^{M}
 + C^{\alpha \beta}\delta^{A}_{\alpha}\delta^{B}_{\beta}\nonumber
\end{align}

 which are equivalent to eqs. (\ref{56}). Direct check shows that the commutation relations of $T^{A}_{\  B}$ matrix
elements implied by eq.(\ref{51}), with $\tilde{\mathcal{B}}^{AB}_{\  \ MN}$ matrix given by eq.(\ref{57}), are not consistent in the sense 
described above (see remark a) bellow the eq.(\ref{53b})). Nevertheless, referring to "gauge" symmetry of $\tilde{\mathcal{B}}^{AB}_{\  \ MN}$ 
matrix (see remark b) one can write RTT relations (eq.(\ref{51})) for the matrix

\begin{align*}
\mathcal{B}^{AB}_{\  \ MN} = \tilde{\mathcal{B}}^{AB}_{\  \ MN} + b^{AB}_{\  \ MN},
\end{align*}

where $b^{AB}_{\  \ MN}$ verifies condition (\ref{55}) and reads:\\

\begin{align}\label{59}
b^{AB}_{\  \ MN} =&\Sigma^{mn}_{\ \dot{\alpha} \dot{\beta}}\delta^{A}_{m}\delta^{B}_{n}\delta^{\dot{\alpha}}_{M}\delta^{\dot{\beta}}_{N}+\frac{i}{2}((\Sigma_{1})^{\alpha m}_{\  \dot{\alpha}}\delta^{A}_{m}\delta^{B}_{\alpha}\nonumber \\& - (\Sigma_{2})^{\alpha m }_{\  \dot{\alpha}}\delta^{A}_{\alpha}\delta^{B}_{m})(\delta^{\dot{\alpha}}_{M}\delta^{4}_{N}-\delta^{4}_{M}\delta^{\dot{\alpha}}_{N})
\end{align}

with $(\Sigma_{1})^{\alpha m}_{\  \dot{\alpha}}$,  $(\Sigma_{2})^{\alpha m }_{\  \dot{\alpha}}$ and $\Sigma^{mn}_{\dot{\alpha}\dot{\beta}}=\Sigma^{mn}_{\dot{\beta}\dot{\alpha}}$ being arbitrary constant matrices.\\
Now, requiring the commutation rules for $T^{A}_{\  B}$  elements (implied by RTT relations with the $\mathcal{B}^{AB}_{\  MN}$ matrix ) to be
consistent allow us to find the matrices $\Sigma$ 

\begin{align}\label{60}
&(\Sigma^{mn})_{\ \dot{\alpha} \dot{\beta}}= -C^{\alpha \beta}(\sigma^{m}_{\alpha \dot{\alpha}}\sigma^{n}_{\beta \dot{\beta}}+ 
\sigma^{n}_{\beta \dot{\beta}}\sigma^{m}_{\alpha \dot{\alpha}})\\
&(\Sigma_{1})^{\alpha m}_{\  \dot{\alpha}}=-(\Sigma_{2})^{\alpha m}_{\  \dot{\alpha}}=
-C^{\alpha \beta}\sigma^{m}_{\beta \dot{\alpha}}\nonumber
\end{align}

and consequently $\mathcal{B}^{AB}_{\  MN}$ matrix  

\begin{align}\label{61}
\mathcal{B}^{AB}_{\  \ MN} =(-1)^{|A||B|}\delta^{A}_{N}\delta^{B}_{M} +C^{\alpha \beta}\delta^{A}_{\alpha}\delta^{B}_{\beta}\delta^{4}_{M}\delta^{4}_{N} - C^{\alpha \beta}\sigma^{m}_{\alpha \dot{\alpha}}\sigma^{n}_{\beta \dot{\beta}}\delta^{A}_{m}\delta^{B}_{n}\delta^{\dot{\alpha}}_{N}\delta^{\dot{\beta}}_{M}\\+ iC^{\alpha \beta}(\sigma^{m}_{\alpha \dot{\alpha}}\delta^{A}_{m}\delta^{B}_{\beta}\delta^{4}_{M}\delta^{\dot{\alpha}}_{N}-\delta^{A}_{\beta}\delta^{B}_{m}\delta^{\dot{\alpha}}_{M}\delta^{4}_{N})\nonumber
\end{align}

The matrix (\ref{61}) defines deformed $N = \frac{1}{2}$ Euclidean superspace described by eqs. (\ref{56}), (\ref{58}) 
 and provides consistent commutation relations for $T^{A}_{\   B}$ elements:
 
\begin{align}\label{61a}
&(-1)^{|A|(|B|+|E|) + |E||F|}T^{B}_{\   E}T^{A}_{\   F} - (-1)^{|F||B|}T^{A}_{\   F}T^{B}_{\   E}=\\
&=C^{\alpha \beta}\sigma^{m}_{\alpha \dot{\alpha}}\sigma^{n}_{\beta \dot{\beta}}((-1)^{|E|(|F|+1)}
\delta^{A}_{m}\delta^{B}_{n}T^{\dot{\beta}}_{\   E}T^{\dot{\alpha}}_{\   F}-T^{A}_{\   m}T^{B}_{\   n}\delta^{\dot{\beta}}_{E}
\delta^{\dot{\alpha}}_{F})\nonumber\\
& + iC^{\alpha \beta}\sigma^{m}_{\alpha \dot{\alpha}}(-(-1)^{|E|(|F|+1)}\delta^{A}_{m}
\delta^{B}_{\beta}T^{4}_{\   E}T^{\dot{\alpha}}_{\   F}+T^{A}_{\   m}T^{B}_{\   \beta}\delta^{4}_{E}\delta^{\dot{\alpha}}_{F})\nonumber \\
&+iC^{\alpha \beta}\sigma^{m}_{\alpha \dot{\alpha}}((-1)^{|E||F|}
\delta^{A}_{\beta}\delta^{B}_{m}T^{\dot{\alpha}}_{\   E}T^{4}_{\   F}-(-1)^{|B|}T^{A}_{\   \beta}T^{B}_{\   m}
\delta^{\dot{\alpha}}_{E}\delta^{4}_{F})\nonumber \\
&+ C^{\alpha \beta}(-(-1)^{|E||F|}(\delta^{A}_{\alpha}\delta^{B}_{\beta}T^{4}_{\   E}T^{4}_{\   F}+
(-1)^{(|B|+1)}T^{A}_{\   \alpha}T^{B}_{\   \beta}\delta^{4}_{E}\delta^{4}_{F})\nonumber
\end{align}

In terms of noncommutative counterparts of Euclidean transformations parameters $(\xi, \bar{\xi}, a, \omega)$ these relations read:

\begin{align}\label{21}
\{\hat{\xi}^{\alpha}, \hat{\xi}^{\beta} \}=&C^{\alpha \beta}_{-}\\
\{\hat{\bar{\xi}}^{\dot{\alpha}}, \hat{\bar{\xi}}^{\dot{\beta}} \}=&0\nonumber\\
\{\hat{\xi}^{\alpha}, \hat{\bar{\xi}}^{\dot{\beta}} \}=&0\nonumber \\
[\hat{a}^{m}, \hat{\xi}^{\alpha}]=& iC_{+}^{\alpha \rho}(\sigma^{m})_{\rho \dot{\rho}}\hat{\bar{\xi}}^{\dot{\rho}}\nonumber \\
[\hat{a}^{m}, \hat{\bar{\xi}}^{\dot{\alpha}}]=&0\nonumber \\
[\hat{a}^{m},\hat{a}^{n}] =& \frac{1}{2} C_{-}^{\alpha \beta} (\sigma^{m})_{\alpha \dot{\gamma}}
(\sigma^{n})_{\beta \dot{\delta}}[\hat{\bar{\xi}}^{\dot{\gamma}}, \hat{\bar{\xi}}^{\dot{\delta}}]
\nonumber
\end{align}

where we have defined :

\begin{align}
&C^{\alpha \beta}_{\pm}=
[\delta^{\alpha}_{\gamma}\delta^{\beta}_{\delta}\pm(A^{T})^{\alpha}_{\  \gamma}(A^{T})^{\beta}_{\   \delta}]C^{\gamma \delta}\nonumber
\end{align}.

This result agrees with the one obtained, in another way, in \cite{a44.1}.\\

One can avoid direct ( i.e. by brute force ) checking whether the B matrix found above satisfies Yang - Baxter equation by noting that this matrix can be written in the form

\begin{align}\label{61b}
\mathcal{B}^{AB}_{\  \ MN} =((I \otimes I +C^{\alpha \beta}Q_{\alpha} \otimes Q_{\beta})\sigma)^{AB}_{\  \ MN}
\end{align}

where:
 
\begin{align}\label{61c}
(Q_{\alpha} \otimes Q_{\beta})^{AB}_{\  \ MN} =&(-1)^{|N|(|B|+|N|+1) + |B|}(Q_{\alpha})^{A}_{\  \ M}(Q_{\beta})^{B}_{\  \ N}\\
\sigma^{AB}_{\  \ MN}=&(-1)^{|A||B|}\delta^{A}_{N}\delta^{B}_{M},\nonumber
\end{align}
 
and $Q_{\alpha}$ are given by eq.(\ref{47}).\\
Taking into account that the supertranslation generators  $Q_{\alpha }$ are nilpotent the $\mathcal{B}^{AB}_{\  \ MN}$ matrix can be rewritten as follows:\\

\begin{align}\label{62}
\mathcal{B}^{AB}_{\  \ MN} =(exp\left\{C^{\alpha \beta}Q_{\alpha} \otimes Q_{\beta}\right\}\sigma)^{AB}_{\  \ MN} 
\equiv (\mathcal{R}\sigma)^{AB}_{\  \ MN},
\end{align}

so that

\begin{align}\label{62a}
\mathcal{R} \equiv \Sigma_{\alpha}\mathcal{R}_{\alpha} \otimes \mathcal{R}_{\alpha} =exp\left\{C^{\alpha \beta} Q_{\alpha} \otimes Q_{\beta}\right\}.
\end{align}

It is not difficult to check that the matrix (\ref{62a}) obeys Y-B equation:\\
\begin{align}
\label{63}
\mathcal{R}_{12}\mathcal{R}_{13}\mathcal{R}_{23} = exp\left\{C^{\alpha \beta} Q_{\alpha} \otimes Q_{\beta} \otimes I + Q_{\alpha} \otimes I \otimes Q_{\beta} + I \otimes  Q_{\alpha} \otimes Q_{\beta} \right\}\nonumber \\=\mathcal{R}_{23}\mathcal{R}_{13}\mathcal{R}_{12}
\end{align}

Our next example deals with Euclidean superspace deformation given by the following nontrivial commutators:  
 
\begin{align}\label{64}
&\left[\hat{x}^{m}, \hat{\eta}^{\alpha}\right] = i\Pi^{m \alpha}\\
&\left[\hat{x}^{m}, \hat{x}^{n}\right]=(\Pi^{ m \alpha}(\sigma^{n})_{\alpha \dot{\alpha}} - \Pi^{n \alpha}(\sigma^{m})_{\alpha \dot{\alpha}})\hat{\bar{\eta}}^{\dot{\alpha}}.\nonumber
\end{align}

This deformation has been studied in Ref. \cite{a36}, \cite{a40} and \cite{a44.1}. It can be viewed as a particular case of general Euclidean superspace deformations considered in \cite{a44.1} and corresponding to the following choice of structure constants:

\begin{align}\label{63a}
C^{\alpha \beta} = 0&  &D^{\dot{\alpha} \dot{\beta}} =0& &E^{\alpha \dot{\beta}}=0\\
\Pi^{m \alpha} \neq 0& &\Delta^{m \dot{\alpha}} =0& &\Theta^{mn} =0\nonumber
\end{align}

The matrix $\tilde{\mathcal{B}}^{AB}_{\  \ MN}$ leading to these relations can be taken in the form:\\

\begin{align}\label{65}
&\tilde{\mathcal{B}}^{AB}_{\   MN}=(-1)^{|A||B|}\delta^{A}_{N}\delta^{B}_{M}+i\Pi^{m\alpha}(\delta^{A}_{m}\delta^{B}_{\alpha}-\delta^{A}_{\alpha}\delta^{B}_{m})\delta^{4}_{M}\delta^{4}_{N}\\&+\frac{1}{2}\Pi^{m\alpha}\sigma^{n}_{\alpha \dot{\alpha}}(\delta^{A}_{m}\delta^{B}_{n}-\delta^{A}_{n}\delta^{B}_{m})(\delta^{\dot{\alpha}}_{M}\delta^{4}_{N}+\delta^{4}_{M}\delta^{\dot{\alpha}}_{N}).\nonumber
\end{align}

Indeed, $\hat{z}^{A}$  coordinate commutation relations given by eq.(\ref{50}) with the above $\tilde{\mathcal{B}}^{AB}_{\  \ MN}$ matrix
read:\\

\begin{align}\label{66}
\hat{z}^{A}\hat{z}^{B}-(-1)^{|A||B|}\hat{z}^{B}\hat{z}^{A} = 
&\Pi^{m \alpha}(\sigma^{n})_{\alpha \dot{\alpha}}(\delta^{A}_{m}\delta^{B}_{n}-\delta^{A}_{n}\delta^{B}_{m})
\delta^{\dot{\alpha}}_{M}\hat{z}^{M}\\
&+i\Pi^{m \alpha}(\delta^{A}_{m}\delta^{B}_{\alpha}-\delta^{A}_{\alpha}\delta^{B}_{m})\nonumber
\end{align}

and are equivalent to the relations (\ref{64}). However, as in the first example this ad-hoc chosen $\tilde B$ matrix reproducing correctly
superspace deformation does not provide consistent commutation relations of $ T^{A}_{\   B}$ variables. To cure this one adds to the matrix 
(\ref{65}) the $b^{AB}_{\  \ MN}$ matrix given by equation:\\

\begin{align}\label{67}
b^{AB}_{\  \ MN}=\frac{1}{2}\Sigma^{mn}_{\ \dot{\alpha}}\delta^{A}_{m}\delta^{B}_{n}(\delta^{\dot{\alpha}}_{M}\delta^{4}_{N}
-\delta^{4}_{M}\delta^{\dot{\alpha}}_{N}),
\end{align}

where $\Sigma^{mn}_{\ \dot{\alpha}}$  are arbitrary constants.\\ 

This modifies commmutation relations of $T^{A}_{\   B}$ elements leaving unchanged those of $z^{A}$ coordinates because $b^{AB}_{\  \ MN}$ matrix
satisfies eq. (\ref{55}).
Now, requiring $\mathcal{B}^{AB}_{\   MN}=\tilde{\mathcal{B}}^{AB}_{\   MN}+b^{AB}_{\  \ MN}$ matrix to yield consistent commutation relations
(\ref{51}) allows us to find the constants $\Sigma $:

\begin{align}\label{68}
\Sigma^{mn}_{\ \dot{\alpha}} = \Pi^{m \alpha}\sigma^{n}_{\alpha \dot{\alpha}} + \Pi^{n \alpha}\sigma^{m}_{\alpha \dot{\alpha}}
\end{align}

and then the explicit form of $\mathcal{B}$:\\

\begin{align}\label{69}
\mathcal{B}^{AB}_{\   MN}=&(-1)^{|A||B|}\delta^{A}_{N}\delta^{B}_{M}+i\Pi^{m\alpha}(\delta^{A}_{m}\delta^{B}_{\alpha}-\delta^{A}_{\alpha}\delta^{B}_{m})\delta^{4}_{M}\delta^{4}_{N}\\&+\Pi^{m\alpha}\sigma^{n}_{\alpha \dot{\alpha}}(\delta^{A}_{m}\delta^{B}_{n}\delta^{\dot{\alpha}}_{M}\delta^{4}_{N}  -\delta^{A}_{n}\delta^{B}_{m}\delta^{4}_{M}\delta^{\dot{\alpha}}_{N}).\nonumber
\end{align}

Consistent commutation relations of $T^{A}_{\   B}$ elements given by eq. (\ref{51}) read now:\\

\begin{align} \label{70}
&(-1)^{|A|(|B|+|E|) + |E||F|}T^{B}_{\   E}T^{A}_{\   F} - (-1)^{|F||B|}T^{A}_{\   F}T^{B}_{\   E}=\\
&=i\Pi^{m\alpha}(T^{A}_{\   m}T^{B}_{\   \alpha}\delta^{4}_{E}\delta^{4}_{F}
-(-1)^{|E||F|}\delta^{A}_{m}\delta^{B}_{\alpha}T^{4}_{\   E}T^{4}_{\   F})\nonumber \\
&+i\Pi^{m\alpha}((-1)^{|B|+1}T^{A}_{\   \alpha}T^{B}_{\   m}\delta^{4}_{E}\delta^{4}_{F}
+(-1)^{|E||F|}\delta^{A}_{\alpha}\delta^{B}_{m}T^{4}_{\   E}T^{4}_{\   F})\nonumber \\
&+i\Pi^{m\alpha}\sigma^{n}_{\alpha \dot{\alpha}}(T^{A}_{\   m}T^{B}_{\   n}\delta^{\dot{\alpha}}_{E}\delta^{4}_{F}
-(-1)^{|E||F|}\delta^{A}_{m}\delta^{B}_{n}T^{\dot{\alpha}}_{\   E}T^{4}_{\   F})\nonumber \\
&+i\Pi^{m\alpha}\sigma^{n}_{\alpha \dot{\alpha}}(-T^{A}_{\   n}T^{B}_{\   m}\delta^{4}_{E}\delta^{\dot{\alpha}}_{F}
+(-1)^{|E|(|F|+1)}\delta^{A}_{n}\delta^{B}_{m}T^{4}_{\   E}T^{\dot{\alpha}}_{\   F})\nonumber
\end{align}\\

or, explicitly in terms of noncommutative counterparts of Euclidean supergroup parameters:

\begin{align}\label{70a}
\{\hat{\xi}^{\alpha}, \hat{\xi}^{\beta} \}=&0\nonumber\\
\{\hat{\bar{\xi}}^{\dot{\alpha}}, \hat{\bar{\xi}}^{\dot{\beta}} \}=&0\nonumber\\
\{\hat{\xi}^{\alpha}, \hat{\bar{\xi}}^{\dot{\beta}} \}=&0\nonumber \\
[\hat{a}^{m}, \hat{\xi}^{\alpha}]=&i\Pi^{m \alpha}_{-}\nonumber \\
[\hat{a}^{m}, \hat{\bar{\xi}}^{\dot{\alpha}}]=&0\nonumber \\
[\hat{a}^{m},\hat{a}^{n}] =& (\Pi_{+}^{m \alpha}(\sigma^{n})_{\alpha \dot{\sigma}} - \Pi^{n \alpha}_{+}(\sigma^{m})_{\alpha \dot{\sigma}})\hat{\bar{\xi}}^{\dot{\sigma}} \nonumber\\
\end{align}

\begin{align}
&\Pi^{m \alpha}_{\pm}=(\delta^{m}_{p}\delta^{\alpha}_{\beta} \pm(e^{-\hat{\omega}})^{m}_{\   p}(A^{T})^{\alpha}_{\    \beta})\Pi^{p \beta}\nonumber\
\end{align}

The same relations have been obtained in \cite{a44.1}.

As in the previous example  $\mathcal{B}^{AB}_{\   MN}$ matrix can be rewritten in terms of generators $Q_{\alpha}$ and $P_{m}$ (given by eq. (\ref{47})):

\begin{align}\label{71a} 
(\mathcal{B})^{AB}_{\   MN}=(I \otimes I + 
i\Pi^{m\alpha}(P_{m} \otimes Q_{\alpha}-	Q_{\alpha} \otimes P_{m}))\sigma)^{AB}_{\   MN} =\\
	(\exp \left\{i\Pi^{m \alpha}[P_{m} \otimes Q_{\alpha}-	
Q_{\alpha} \otimes P_{m}]\right\}\sigma)^{AB}_{\   MN}\equiv (\mathcal{R}\sigma)^{AB}_{\   MN} \nonumber
\end{align}

where\\

\begin{align}\label{71b}  
(Q_{\alpha} \otimes P_{m})^{AB}_{\   MN} = (-1)^{|M|(|B|+|N|) +|B|}(Q_{\alpha})^{A}_{\   M}(P_{m})^{B}_{\   N}.
\end{align}

Therefore, the R - matrix reads

\begin{align}\label{72}
\mathcal{R}=\exp \left\{i\Pi^{m \alpha}(P_{m} \otimes Q_{\alpha}-	Q_{\alpha}\otimes P_{m})\right\}
\end{align}

This form of  $\mathcal{B}$ matrix is particularly convenient for checking the validity of Yang-Baxter equation.\\

The approach illustrated in the above two examples can be applied to the case of Euclidean version of the so called $\Theta $-Minkowski 
space studied, for example, in Refs, \cite{a42}, \cite{a43}. It is defined by the commutators

\begin{align}\label{72a}
[\hat{x}^m , \hat{x}^n ] = i\Theta_{mn}
\end{align}

with antisymmetric $\Theta^{mn}$ matrix.\\
These commutation relations, as well as the algebraic sector of their symmetry transformations, are controlled by the matrix

\begin{align}\label{72b}
\mathcal{B}^{AB}_{\  \ MN} = 
(\exp \left\{i\Theta^{mn}P_{m} \otimes P_{n}\right\}\sigma)^{AB}_{\  \ MN} \equiv (\mathcal{R}\sigma)^{AB}_{\  \ MN}
\end{align}

Actually, all three matrices given by eqs.(\ref{62}), (\ref{71a}) and (\ref{72b}) can be combined into the following one      

\begin{align}\label{73a}
\mathcal{B}^{AB}_{\   MN}&=(\exp \{- i\Theta^{mn}P_{m} \otimes P_{n}+ C^{\alpha \beta}Q_{\alpha} \otimes Q_{\beta}\\
& + i\Pi^{m \alpha}(P_{m} \otimes Q_{\alpha}-	Q_{\alpha}\otimes P_{m})\}\sigma)^{AB}_{\   MN} \equiv (\mathcal{R}\sigma)^{AB}_{\   MN}\nonumber
\end{align}

This matrix verifies all three conditions (a - c) and describes three parameter extension of $ N = \frac{1}{2}$ Euclidean supersymmetry
considered in \cite{a36}. The relevant commutation relations of space-time coordinates read

\begin{align}\label{72c}
\{\hat{\eta}^{\alpha}, \hat{\eta}^{\beta}\}=& C^{\alpha \beta}\\
[\hat{x}^{m}, \hat{\eta}^{\alpha}]=&iC^{\alpha \beta}(\sigma^{m})_{\beta \dot{\rho}}\hat{\bar{\eta}}^{\dot{\rho}}+ i\Pi^{m \alpha} \nonumber \\
[\hat{x}^{m},\hat{x}^{n}] = &\frac{1}{2} C^{\alpha \beta} (\sigma^{m})_{\alpha \dot{\gamma}}(\sigma^{n})_{\beta \dot{\delta}}[\hat{\bar{\eta}}^{\dot{\gamma}}, \hat{\bar{\eta}}^{\dot{\delta}}]\nonumber \\& + (\Pi^{m \alpha}(\sigma^{n})_{\alpha \dot{\sigma}} - \Pi^{n \alpha}(\sigma^{m})_{\alpha \dot{\sigma}})\hat{\bar{\eta}}^{\dot{\sigma}} + i\Theta^{mn}\nonumber
\end{align}

The algebraic sector of global transformations preserving (\ref{72c}) is given by the following commutators of transformation parameters:
 
\begin{align}\label{72d}
\{\hat{\xi}^{\alpha}, \hat{\xi}^{\beta} \}=&C^{\alpha \beta}_{-}\\
\{\hat{\bar{\xi}}^{\dot{\alpha}}, \hat{\bar{\xi}}^{\dot{\beta}} \}=&0\nonumber\\
\{\hat{\xi}^{\alpha}, \hat{\bar{\xi}}^{\dot{\beta}} \}=&0\nonumber \\
[\hat{a}^{m}, \hat{\xi}^{\alpha}]=&iC_{+}^{\alpha \rho}(\sigma^{m})_{\rho \dot{\rho}}\hat{\bar{\xi}}^{\dot{\rho}}+ i\Pi^{m \alpha}_{-}\nonumber \\
[\hat{a}^{m}, \hat{\bar{\xi}}^{\dot{\alpha}}]=&0\nonumber \\
[\hat{a}^{m},\hat{a}^{n}] =& \frac{1}{2} C_{-}^{\alpha \beta} (\sigma^{m})_{\alpha \dot{\gamma}}
(\sigma^{n})_{\beta \dot{\delta}}[\hat{\bar{\xi}}^{\dot{\gamma}}, \hat{\bar{\xi}}^{\dot{\delta}}]\nonumber \\
& + (\Pi_{+}^{m \alpha}(\sigma^{n})_{\alpha \dot{\sigma}} - \Pi^{n \alpha}_{+}(\sigma^{m})_{\alpha \dot{\sigma}})\hat{\bar{\xi}}^{\dot{\sigma}} + i\Theta^{mn}_{-}\nonumber
\end{align}
where $C^{\alpha \beta}_{\pm}$ etc. are defined as follows

\begin{align}\label{72e}
&C^{\alpha \beta}_{\pm}=[\delta^{\alpha}_{\gamma}\delta^{\beta}_{\delta}\pm(A^{T})^{\alpha}_{\  \gamma}(A^{T})^{\beta}_{\   \delta}]C^{\gamma \delta}\\
&\Pi^{m \alpha}_{\pm}=(\delta^{m}_{p}\delta^{\alpha}_{\beta} \pm(e^{-\hat{\omega}})^{m}_{\   p}(A^{T})^{\alpha}_{\    \beta})\Pi^{p \beta}\nonumber\\
&\Theta^{mn}_{-} = (\delta^{m}_{p}\delta^{n}_{q}-(e^{-\hat{\omega}})^{m}_{\  p}(e^{-\hat{\omega}})^{n}_{\  q})\Theta^{pq}\nonumber
\end{align}

Finally, let us note that the general Euclidean superspace deformations considered in \cite{a44.1} can also be described by a sort of B matrix
constructed out of translations $P_{m}$ and supertranslations $Q_{\alpha}$, $\bar{Q}_{\dot{\alpha}}$ 

\begin{align}\label{74} 
&\mathcal{B}^{AB}_{\   MN}=(I \otimes I -i\Omega^{mn}P_{m} \otimes P_{n} +C^{\alpha \beta} Q_{\alpha} \otimes Q_{\beta}\\& + i\Pi^{m \alpha}(P_{m} \otimes Q_{\alpha}-	Q_{\alpha}\otimes P_{m})\nonumber \\& + D^{\dot{\alpha} \dot{\beta}}\bar{Q}_{\dot{\alpha}} \otimes \bar{Q}_{\dot{\beta}} +i\Delta^{m \dot{\alpha}}(P_{m} \otimes \bar{Q}_{\dot{\alpha}}-	\bar{Q}_{\dot{\alpha}}\otimes P_{m})\nonumber \\& -E^{\alpha \dot{\beta}}(Q_{\alpha} \otimes \bar{Q}_{\dot{\beta}} + \bar{Q}_{\dot{\beta}} \otimes Q_{\alpha}))\sigma)^{AB}_{\   MN}\nonumber
\end{align}

with

\begin{align}\label{74a}
\Omega^{mn} &\equiv \Theta^{mn} -iC^{\alpha \beta}D^{\dot{\alpha} \dot{\beta}}(\sigma^{m})_{\alpha \dot{\alpha}}(\sigma^{n})_{\beta \dot{\beta}}\\
& -\frac{i}{2}E^{\alpha \dot{\beta}}E^{\beta \dot{\alpha}}((\sigma^{m})_{\alpha \dot{\alpha}}(\sigma^{n})_{\beta \dot{\beta}} + 
(\sigma^{n})_{\alpha \dot{\alpha}}(\sigma^{m})_{\beta \dot{\beta}})\nonumber
\end{align}

and  C, D, E, $\Pi $, $\Delta $ being arbitrary constants of properly defined parity.\\
However, without imposing some additional and rather strong constraints on these constants the above B matrix does not satisfy Yang-Baxter equation.\\

In order to complete our discussion of global symmmetry transformations of $N = 1$ quantized Euclidean superspaces we introduce the coalgebraic structure in the set of these transformations. It is defined in the standard way by the following coproduct $\Delta $, counit $\epsilon $ and 
antipode $S$ maps:

\begin{align}\label{75}
\Delta(T^{A}_{\  B}) = T^{A}_{\  C} \otimes T^{C}_{\   B}\
\end{align}
 
\begin{align}\label{76}
\varepsilon(T^{A}_{\    B}) = \delta^{A}_{B}
\end{align}

\begin{align}\label{77}
S(T^{A}_{\   B}) = (T^{-1})^{A}_{\    B}
\end{align}

The explicit form of the inverse $(T^{-1})^{A}_{\    B}$  Euclidean supermatrix reads:\\

\begin{align}\label{78}
(T^{-1})^{A}_{\    B}= \begin{bmatrix}(e ^{\omega})^{m}_{\;\;n}& &-(e ^{\omega})^{m}_{\;\; n}a^{n} & & i\xi^{\alpha}
(e ^{\omega})^{m}_{\;\;n}(\sigma^{n})_{\alpha \dot{\alpha}} & & i\bar{\xi}^{\dot{\alpha}}
(e ^{\omega})^{m}_{\;\;n}(\sigma^{n})_{\alpha \dot{\alpha}} \cr  & &\cr 0 &  & 1 &   &  0 &   &0&  \cr    & 
&\cr 0 &  &-((B^{\dagger})^{-1})^{\dot{\rho}}_{\  \dot{\alpha}}\bar{\xi}^{\dot{\alpha}}&  & 
[(B^{\dagger})^{-1}]^{\dot{\rho}}_{\  \dot{\beta}} &  & 0 \cr \cr 0 & 
&-((A^{T})^{-1})^{\rho}_{\  \alpha}\xi^ {\alpha} & & 0  & & [(A^{T})^{-1}]^{\rho}_{\   \beta}     \cr  \end{bmatrix}
\end{align}\\

One can easily check that the coalgebraic structure introduced in such a way coincides with the one considered in \cite{a44.1}.
What is more the B matrices describing $ N = \frac{1}{2} $ supersymetry and its three-parameter extension satisfy the following
equation:  

\begin{align} 
\Delta((-1)^{|M|(|N|+|B|)} T^{A}_{\  M}T^{B}_{\  N}B^{MN}_{\  EF}) = \Delta((-1)^{|E|(|F|+|N|)}B^{AB}_{\  MN} T^{M}_{\  E}T^{N}_{\  F}).
\end{align}

This proves that the coproduct is compatible with (anti)commutation relations obeyed by transformation parameters. 

\section{Star products, twisted Euclidean superalgebras and global symmetries} 

In this last section we briefly discuss the global symmetries of deformed superspaces considered above and in Ref.\cite{a44.1} from the
point of view of star products and twisted superalgebras \cite{a36}, \cite{a38}, \cite{a55} - 
\cite{a61}. As it is well known noncommutative (super)spaces can be regarded as the usual ones equipped, however, with additional structure 
(the so called star product). This product is defined to be associative but not commutative which allows to model noncommutative (super)spaces.
Twisted (super)algebras appear, in a natural way, when analyzing covariance of the commutation relations defining these (super)spaces.
In particular, the general commutation relations yielding the deformations of Euclidean superspace studied in \cite{a44.1}, as well as their special cases described above (see eqs. (\ref{56}), (\ref{64}) and (\ref{72c}) ), can be realized as commutation relations       
$[x^{A}\stackrel{\star}{,} x^{B}]_{\pm } =x^{A} \star x^{B} \pm  x^{B}\star x^{A}$ corresponding to the star product given by  

\begin{align}
f\star g \equiv \mu \circ \mathcal{F_{\mathcal{E}}}^{-1}(f \otimes g) = \mu_{\star}(f \otimes g).
\end{align}
In the above formula, $\mu $ denotes standard multiplication of functions on Euclidean superspace  $\mu(f \otimes g) = fg$
while $\mathcal{F}_{\mathcal{E}}$ is an
operator constructed out of differential realization of translations $P_{m}$ and supertranslations
$Q_{\alpha}$, $\bar{Q}_{\dot{\alpha}}$ 

\begin{align}\label{79}
\mathcal{F}_{\mathcal{E}} =& exp\{\frac{1}{2}(i\Omega^{mn}P_{m} \otimes P_{n} - C^{\alpha \beta}Q_{\alpha} \otimes Q_{\beta}\\
& +i\Pi^{m \alpha}(Q_{\alpha} \otimes P_{m} - P_{m} \otimes Q_{\alpha}) - 
D^{\dot{\alpha}\dot{\beta}}\bar{Q}_{\dot{\alpha}} \otimes \bar{Q}_{\dot{\beta}}\nonumber \\
& + i\Delta^{m \dot{\alpha}}(\bar{Q}_{\dot{\alpha}} \otimes P_{m} -P_{m} \otimes \bar{Q}_{\dot{\alpha}})\nonumber\\
& + E^{\alpha \dot{\beta}}(Q_{\alpha} \otimes \bar{Q}_{\dot{\beta}}+ \bar{Q}_{\dot{\beta}}\otimes Q_{\alpha})\}\nonumber
\end{align}

\begin{align}\label{80}
Q_{\alpha} =& i \frac{\partial}{\partial \eta^{\alpha}} + \sigma^{m}_{\alpha \dot{\alpha}}\bar{\eta}^{\dot{\alpha}} \partial_{m}\\
\bar{Q}_{\dot{\alpha}} =& -i \frac{\partial}{\partial \bar{\eta}^{\dot{\alpha}}} - \sigma^{m}_{\alpha \dot{\alpha}}\eta^{\alpha} \partial_{m}\nonumber\\
P_{m} =&-i\partial_{m}\nonumber
\end{align}

The deformed Euclidean superspace and its $N = \frac{1}{2}$ supersymmetry considered in the first example correspond to the case where
C is an arbitrary constant symmetric matrix while the remaining constants D, E, $\Pi$ and  $\Delta $ vanish. In the second example it is the constant
$\Pi $ which is the only non-vanishing one. Finally, three-parameter extension of $N = \frac{1}{2}$ supersymmetry described by B matrix given by eq.(\ref{73a}) corresponds
to nonzero C, $\Pi $ and $\Theta $. \\
Now, the star counterparts of general commutation relations generated by the $\mathcal{F}_{\mathcal{E}}$  operator
with arbitrary nontrivial constant matrices C, D, E, $\Pi $, $\Delta $ and $\Theta $ are not consistent with Jacobi identities (unless the elements of
these matrices belong to some Grassmann algebra). However, in the special case of $N = \frac{1}{2}$ supersymmetry and its three-parameter extension the relevant $\mathcal{F}_{\mathcal{E}}$ operator does satisfy the conditions defining supersymmetric extension of  Drinfeld twist operator \cite{a22}, \cite{a32}, \cite{a36}, \cite{a38}. This implies that:
\begin{enumerate}[i)]
\item star product given by the $\mathcal{F}_{\mathcal{E}}$  operator is associative and commutation relations corresponding to this product
are consistent with Jacobi identities,

\item one can construct the so called universal $\mathcal{R}$ matrix \cite{a22}, \cite{a25},  according to the general rule   
 
\begin{align}\label{82}
\mathcal{R} = {(\mathcal{F}_{\mathcal{E}})}_{21}{\mathcal{F}_{\mathcal{E}}}^{-1},
\end{align}

where\\

\begin{align}
{(\mathcal{F}_{\mathcal{E}})}_{21}=\sigma({\mathcal{F}_{\mathcal{E}}})\nonumber
\end{align}

and $\sigma $ is a flip operator.

$\mathcal{R}$ matrix, constructed in such a way, when expressed in terms of the generators $P_{m}$, $Q_{\alpha}$, 
$\bar{Q}_{\dot{\alpha}}$, reads 
       
\begin{align}\label{83}
\mathcal{R}=&exp\{-(i\Theta^{mn}P_{m}\otimes P_{n} - C^{\alpha \beta}Q_{\alpha}\otimes Q_{\beta}\\
&+i\Pi^{m\alpha}(Q_{\alpha}\otimes P_{m}-P_{m}\otimes Q_{\alpha}))\}\nonumber
\end{align} 

This shows that the B supermatrix (\ref{73a}) is (up to the flip operator ) nothing but the supermatrix realization of universal $\mathcal{R}$ matrix given by eq. (\ref{83}). 
Obviously, the same conclusion holds in the special cases when only one of the constants $\Theta  $, $C$, $\Pi $ does not vanish.
    
\item star counterparts of commutation relations are covariant with respect to the action of:

a) twisted $N = \frac{1}{2}$ Euclidean superalgebras. It means that for any generator $X$ of this superalgebra the  following equation holds

\begin{align}\label{84}
X\mu_{\star}(f \otimes g) = \mu_{\star}(\Delta_{\mathcal{F}}(X)(f \otimes g)).
\end{align}

\begin{align}\label{86}
\Delta_{\mathcal{F}}(x) =  \mathcal{F}( X \otimes I + I \otimes X)\mathcal{F}^{-1}.
\end{align}

$\Delta_{\mathcal{F}}(x)$ is the twisted coproduct.

b) global $N = \frac{1}{2}$ Euclidean supersymmetry transformations with noncommutative parameters satisfying commutation relations described within FRT method by eq.(\ref{51}) (with B matrix given by eq. (\ref{73a})). These relations (given explicitely by eq.(\ref{72d})) have also been obtained, in independent way, in \cite{a44.1}.            
\end{enumerate}

\section{Summary }

A (super)space quantization breaking, in general, the usual (super)symmetry calls for a modification of symmetry transformations. At infinitesimal level this can be achieved in the framework of twisted (super)algebras. The twist operator modifies coalgebraic sector of transformations leaving the algebraic one unchanged. In the dual, global picture, symmetry transformations act on deformed superspaces in the usual way but they depend on noncommutative parameters. The most interesting cases correspond to the situation when these parameters generate quantum symmetry supergroup ( noncommutative Hopf superalgebra). In particular, this happens for $ N = \frac{1}{2}$ supersymmetry and its extensions studied in many papers. At infinitesimal level these deformations are preserved by appropriately twisted superalgebras \cite{a32}, \cite{a36}, \cite{a38}, \cite{a39}. The gobal symmetry transformations preserving them have been studied in \cite{a44.1} (see also Ref. \cite{a32}) and shown to generate supersymmetric extensions of $\theta $-Euclidean group.

In the present paper we have demonstrated how these extensions can be obtained within the FRT method. Using supervector realization of (deformed) Euclidean superspaces and supermatrix representation $T^{A}_{B}$ of Euclidean supergroup we have found explicit form of R-matrices corresponding to
$N = \frac{1}{2}$ supersymmetry and its three-parameter generalization. These matrices control both coordinate commutation relations defining deformed superspace as well as commutators of parameters defining algebraic sector of their global symmetry transformations. 
We have shown that such matrices can be constructed directly, starting, basically, with coordinate commutators (in particular, without referring to the twisting  procedure). Actually, superspace deformations do not determine the corresponding R-matrices uniquely. However, as we demonstrated, in our cases this sort of gauge freedom can be used (due to a specific structure of some $T^{A}_{B}$ matrix elements of Euclidean supergroup representation) to impose the consistent supersymmetric version of FRT equations. Explicit form of supermatrix representation of Euclidean supergroup allowed us to find commutators of transformation parameters as well as to introduce, in a standard manner, the coalgebraic structure (compatible with the algebraic one) in the set of global transformations preserving $ N = \frac{1}{2}$ supersymmetry and its extensions. In this way, we confirmed independently the results of \cite{a44.1} concerning quantum supergroup structure of global symmetries.\\ 
In the last section of the paper the global transformations have been analyzed in the star products and twisted superalgebras perspective. In particular, we argued that R matrices found in the first sections provide supermatrix realization of universal R-matrices constructed out of relevant twist operators which define twisted Euclidean superalgebras preserving $ N = \frac{1}{2}$ supersymmetry.                  

\section{Acknowledgment}

We thank Piotr Kosiński and Paweł Maślanka for fruitful discussion and reading the manuscript.\\
The paper is supported by the grant of he Ministry of Science N N202 331139 and Lodz University grant 506/1037.\\

\section{Appendix; Supermatrices}

Bellow we collect some basic facts concerning supermatrices \cite{a63} 

A matrix S is a supermatrix if its elements $S_{ij}$ belong to some superalgebra and its rows and columns  have definite parities. If
$|S_{ij}| = 0,1 $ denotes the parity of $ S_{ij}$ element, $|i|, |j| = 0,1$ - the parity of i-th row and j-th column j, the parity $ |S| = 0,1 $ of the supermatrix S is defined by 

\begin{align}
&|S| = 0, \mbox{ \,-\,( $S$ is an even supermatrix ) if: \,\ } |S_{ij}|+|i|+|j| =\mbox{ 0 mod 2, and} \\
&|S| = 1, \mbox{ \,-\,( $S$ is an odd supermatrix ) if \,\ } |S_{ij}|+|i|+|j| =\mbox{1 mod 2}\nonumber \\
&\mbox{for all \,\ } i,j.\nonumber
\end{align}  

Usually even rows and columns are put first. Then a supermatrix S which has m even, n odd rows and p even, q odd columns can be put in the form

\begin{align}
S = \begin{bmatrix}S_{00}& &S_{01}\\ S_{10}&  &S_{11}  \end{bmatrix}
\end{align}
where $S_{00}$, $S_{01}$, $S_{10}$, $S_{10}$ denote  $m \times p$, $n \times p$, $m \times q$, $n \times p$ matrices respectively. According to the definition given above 

\begin{align*}
|S| = 0   \mbox{\,\,if\,\,} &|(S_{00})_{ij}| = |(S_{11})_{ij}| =0\\
&|(S_{01})_{ij}| = |(S_{01})_{ij}| =1\\
\mbox{\,\,and\,\,}\\
|S| = 1   \mbox{\,\,if\,\,} &|(S_{00})_{ij}| = |(S_{11})_{ij}|=1\\
&|(S_{01})_{ij}| = |(S_{01})_{ij}| =0
\end{align*}

The sum and product of the supermatrices ( which can be divided into blocks of compatible dimensions ) are defined as in the case of ordinary matrices. However, the definition of  multiplication of supermatrices by scalars has to take into account that there are odd elements in superalgebra. Therefore, the left- and right-hand side multiplication of supermatrices  by scalar $\alpha $ are defined respectively, as follows

\begin{align}\label{d1}
\alpha \cdot S =  \begin{bmatrix}\alpha S_{00}& &\alpha S_{01}\\ (-1)^{|\alpha}|\alpha S_{10}&  &(-1)^{|\alpha|}\alpha S_{11}  \end{bmatrix}
\end{align}

\begin{align}\label{d2}
S \cdot \alpha =  \begin{bmatrix} S_{00}\alpha& &S_{01}\alpha\\  S_{10}(-1)^{|\alpha}|\alpha&  &S_{11}(-1)^{|\alpha|}\alpha   \end{bmatrix}
\end{align}

Matrix elements $(S_{1} \otimes S_{2})^{AB}_{\  \ MN}$ of the tensor product $S_{1} \otimes S_{2}$ are defined according to general rule of multiplication in the tensor product of superalgebras i.e.

\begin{align}
((S_{1} \otimes S_{2}))((S'_{1} \otimes S'_{2})) = (-1)^{|S_{2}||S'_{1}|}S_{1}S'_{1} \otimes S_{2}S'_{2}
\end{align}  

where $S_{1}, S'_{1}, S_{2}, S'_{2}$ denote supermatrices for which the products $S_{1}S'_{1}$ and $S_{2}S'_{2}$ are well defined. Hence, for even  $P$ ($|P| = 0$) and odd $N$ ($|N|=1$) supermatrices the relevant  matrix elements of their tensor products read: 
 
\begin{align}
&(P_{1} \otimes P_{2})^{AB}_{\,\ CD} = (-1)^{|C|(|B|+|D|)}(P_{1})^{A}_{\,\ C}(P_{2})^{B}_{\,\ D}\\
&(N_{1} \otimes N_{2})^{AB}_{\,\ CD} = (-1)^{|C|(|B|+|D|+1)+|B|}(N_{1})^{A}_{\,\ C}(N_{2})^{B}_{\,\ D}\nonumber \\
&(N \otimes P)^{AB}_{\,\ CD} = (-1)^{|C|(|B|+|D|)+|B|}N^{A}_{\,\ C}P^{B}_{\,\ D}\nonumber \\
&(P \otimes N)^{AB}_{\,\ CD} = (-1)^{|C|(|B|+|D|+1)}P^{A}_{\,\ C}N^{B}_{\,\ D}\nonumber 
\end{align}

\end{document}